\documentclass{article}

\PassOptionsToPackage{numbers, compress}{natbib}

\usepackage[final]{neurips_2021}

\usepackage[utf8]{inputenc} 
\usepackage[T1]{fontenc}    
\usepackage{hyperref}       
\usepackage{url}            
\usepackage{booktabs}       
\usepackage{amsfonts}       
\usepackage{nicefrac}       
\usepackage{microtype}      
\usepackage{xcolor}         
\usepackage{graphicx}

\title{Generalised DePIN Protocol: A Framework for Decentralized Physical Infrastructure Networks}

\author{Dipankar Sarkar \\
  Cryptuon Research \\
  \texttt{me@dipankar.name} \\
}

\begin{document}

\maketitle

\begin{abstract}
This paper introduces the Generalised DePIN (GDP) protocol, a comprehensive framework for decentralized physical infrastructure networks. GDP establishes a modular system, enabling tailored application across sectors like ridesharing and power systems. Leveraging device onboarding, multi-sensor redundancy, and a reward/penalty mechanism, GDP promotes genuine behavior and ensures network-wide vigilance. Through continuous audits and updates, the protocol remains dynamic, ensuring sustainable decentralized operations.
\end{abstract}

\section{Introduction}

The evolution of decentralized systems has been rapid, affecting various sectors of the economy and society \cite{nakamoto2008bitcoin} . Such systems promise increased transparency, reduced intermediaries, and enhanced user control, but also bring challenges in terms of security, scalability, and trustworthiness \cite{buterin2015ethereum}. As these systems interact increasingly with the physical world, there's a pressing need for a robust framework that addresses these concerns. Enter the Generalised DePIN (Decentralized Physical Infrastructure Networks) Protocol, or GDP.

GDP is more than just a protocol; it's a holistic framework designed to be adaptable across various real-world applications. Whether it's a decentralized ridesharing service, a peer-to-peer power distribution system, or any other infrastructure that can benefit from a decentralized approach, GDP provides the foundational elements required for such systems to function efficiently and securely \cite{tapscott2016blockchain}.

The cornerstone of GDP is its emphasis on initialization. Proper onboarding of devices or participants is crucial to ensure that only legitimate entities participate in the network. This is achieved through advanced cryptographic techniques, such as Zero-Knowledge Proofs (ZKPs) and Multi-Party Computation (MPC), which offer a high level of security without compromising on privacy \cite{bensasson2013snarks}. Furthermore, the stake deposit mechanism acts as a deterrent against malicious activities, ensuring that participants have a vested interest in the network's wellbeing.

Operational robustness is another hallmark of the GDP. By employing multi-sensor redundancy and peer witness systems, the protocol ensures that critical actions are validated from multiple angles, reducing the chance of false data injection or malicious activities. The commit-reveal mechanism and stochastic checks further enhance the operational integrity of the system, ensuring that participants remain honest and genuine in their dealing \cite{wood2014ethereum}.

However, even with the best of intentions, discrepancies can arise. This is where GDP's validation and resolution mechanisms come into play. Using advanced machine learning models, the system can detect anomalies and potentially malicious activities. The community oversight feature further reinforces this by enabling participants to report suspicious activities, ensuring that the network remains vigilant at all times\cite{zohar2015bitcoin}.

GDP also recognizes the importance of continual improvement. Through periodic audits and updates, the protocol remains dynamic, adapting to the latest technological advancements and feedback from its community. This iterative approach ensures that GDP remains at the forefront of decentralized system frameworks, paving the way for more secure, transparent, and efficient decentralized applications in the future.

In this paper, we will delve deeper into each component of the GDP, providing insights into its design, functionality, and potential applications. Through real-world examples, we will demonstrate how GDP can revolutionize various sectors, making them more resilient and user-centric.

\section{Existing works}

Decentralized Physical Infrastructure Networks (DePINs) have garnered significant interest in recent years, promising to revolutionize various sectors by marrying the physical world with decentralized protocols. Several DePINs have emerged, each with its unique functionalities and challenges. 

While existing DePINs have made remarkable strides in decentralizing physical infrastructures, challenges remain. Scalability, security, potential centralization, and data verification are recurrent themes. It's within this context that our proposed Generalised DePIN (GDP) protocol emerges, aiming to address these drawbacks and push the boundaries of what DePINs can achieve.

Here, we discuss notable DePINs, shedding light on their features and identified shortcomings.

\subsection{IoTeX Network}
\textbf{Functionality} IoTeX \cite{raftopoulos2019iotex} is primarily focused on connecting the Internet of Things (IoT) devices in a decentralized manner. The network aims to ensure scalability, privacy, and interoperability between connected devices.

\textbf{Drawbacks} While IoTeX offers a promising solution for IoT devices, concerns have been raised about its long-term scalability, especially as the number of IoT devices worldwide is expected to surge. Additionally, while it emphasizes privacy, certain edge cases might still expose user data.

\subsection{Helium Network}
\textbf{Functionality} Helium \cite{hale2020helium} takes a unique approach by allowing users to deploy wireless coverage (LoRaWAN) in exchange for cryptocurrency. It has rapidly expanded its decentralized wireless infrastructure across many urban areas.

\textbf{Drawbacks} The primary concern with Helium is its reliance on a single type of hardware, leading to potential centralization issues. Additionally, security challenges related to the initial setup of hotspots have been reported.

\subsection{IOTA}
\textbf{Functionality} IOTA \cite{popov2018tangle}, unlike traditional blockchains, uses a directed acyclic graph (DAG) called the Tangle. It's designed for micro-transactions between IoT devices without incurring fees.

\textbf{Drawbacks} Concerns have arisen regarding IOTA's centralization due to the Coordinator, a node run by the IOTA Foundation to prevent attacks. Though plans are in place to remove it, the transition remains a significant challenge. Additionally, concerns about the network's susceptibility to quantum attacks have been voiced.

\subsection{Streamr Network}
\textbf{Functionality} Streamr \cite{patel2021streamr} aims to create a decentralized marketplace for real-time data. It focuses on data from IoT devices, social media platforms, and other online sources.

\textbf{Drawbacks} Streamr is still in its nascent stages, and concerns about data quality and verification persist. Ensuring that data providers offer genuine, unaltered data remains a significant challenge.

\subsection{MXC Foundation's Data Network}
\textbf{Functionality} This project focuses on creating a global data network to facilitate the transfer and trade of wireless data. It combines blockchain technology with LPWAN (Low Power Wide Area Network) technology.

\textbf{Drawbacks} Like Helium, MXC \cite{werner2019mxc} relies heavily on specific hardware components, potentially leading to centralization. Also, its global adoption rate remains a point of contention.

\section{Generalised DePIN Protocol : Introduction}

In today's age of rapid technological advancement, the line between the physical and digital realms is increasingly blurred. With the rise of the Internet of Things (IoT) and a myriad of smart devices permeating our daily lives, the need for robust, secure, and efficient Decentralized Physical Infrastructure Networks (DePINs) has never been greater. 

\textbf{The underpinning challenge? }Ensuring that devices participating in these networks are genuine, verified, and trustworthy. Enter software-driven techniques for device verification—a central theme of our proposed Generalised DePIN (GDP) protocol.

\subsection{Key challenges}
At the heart of any DePIN lies the devices that constitute its network. Ensuring the authenticity and integrity of these devices is paramount. Traditional methods, often hardware-based, have shown vulnerabilities to physical tampering and external attacks \cite{chen2017hardware}. 

Software-driven techniques offer a more dynamic approach, leveraging the power of advanced algorithms, cryptographic methods, and real-time data processing. For instance, Zero-Knowledge Proofs (ZKPs) and Multi-Party Computation (MPC) can validate device credentials without revealing sensitive information, offering both security and privacy \cite{goldwasser1989knowledge}.

\subsubsection{Hardware Dependence Dilemma}
Traditionally, many systems have relied on hardware attestations for device onboarding. This approach leans heavily on trust in the device manufacturer, assuming that if the hardware is genuine and untampered, then the device can be trusted. However, several challenges arise:

\textbf{Supply Chain Attacks} Recent years have seen a surge in sophisticated supply chain attacks where malicious actors compromise hardware components during the manufacturing or distribution process\cite{perlroth2020supply}.

\textbf{Hardware Obsolescence}Hardware components age and can become obsolete, potentially introducing vulnerabilities if not regularly updated or replaced.

\textbf{Limited Flexibility} Hardware-based solutions often lack the flexibility to adapt to new threats or to be updated with the latest security protocols.

\subsubsection{Software-Centric Onboarding Paradigm}
Given the limitations of hardware dependence, a shift towards software-driven device onboarding is not just preferable—it's imperative. Here's how a software-centric approach addresses the challenges:

\textbf{Dynamic Verification} Software allows for dynamic and continuous verification processes. Techniques such as Zero-Knowledge Proofs (ZKPs) can validate a device's credentials without revealing the actual data, ensuring both security and privacy \cite{bensasson2014succinct}.

\textbf{Adaptability} Software solutions can be updated remotely, ensuring that the latest security protocols are always in place and that the system can adapt to emerging threats.

\textbf{Reduced Trust in Manufacturers} By moving the trust anchor from hardware to software, the system reduces its dependence on device manufacturers, mitigating risks associated with hardware tampering.

\textbf{Cost-Effective} Software-based solutions, especially those leveraging open-source frameworks, can be more cost-effective in the long run, eliminating the need for frequent hardware replacements or upgrades.

\subsection{Protocol components}

For a DePIN to function cohesively and securely, we have identified the following primary process flows. Each of these represent an important part of the steps required to run such a system holistically for the end customers and providers. 

\begin{figure}[ht]
  \centering
  \includegraphics[scale=0.3]{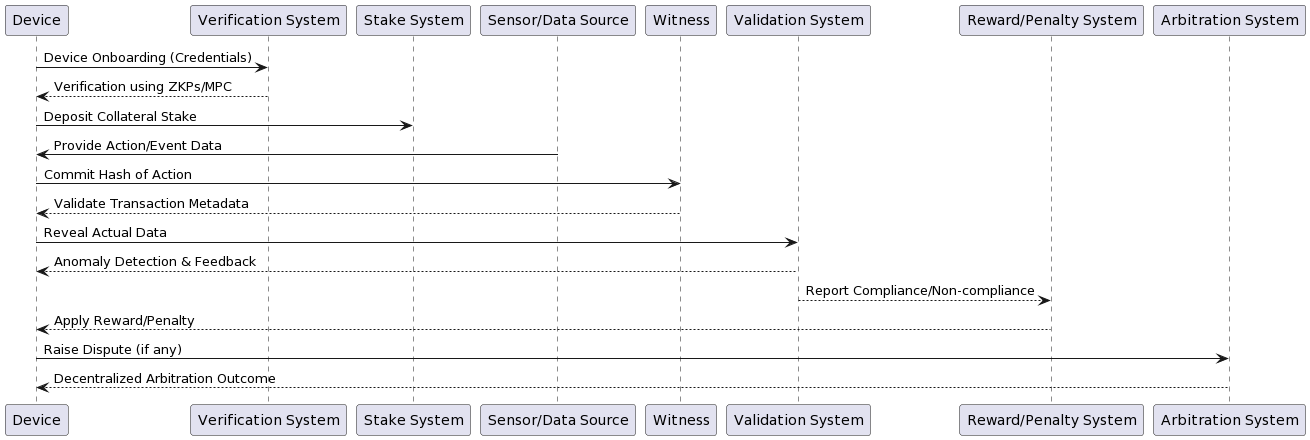}
  \caption{Sequence diagram of components}
\end{figure}

\textbf{Device Onboarding}The initial step where devices join the DePIN. This involves not only the verification of the device's credentials but also its history, ensuring it hasn't been part of malicious activities in other networks.

\textbf{Data Transmission} Once onboarded, devices will continually transmit and receive data. Ensuring the integrity and authenticity of this data flow is vital. Techniques like encryption, timestamping, and digital signatures play a pivotal role here.

\textbf{Consensus Mechanism} In decentralized networks, achieving agreement on data or transaction validity is crucial. Whether it's Proof of Work (PoW), Proof of Stake (PoS), or more energy-efficient algorithms, the consensus mechanism ensures that data is accepted and recorded in a decentralized manner \cite{merkle1987digital}.

\textbf{Anomaly Detection}With the influx of massive data, detecting outliers or malicious patterns is a challenge. Machine learning models can continuously analyze this data flow, flagging potential anomalies for review.

\textbf{Reward/Penalty Mechanisms} To incentivize genuine behavior and deter malicious actors, a system of rewards and penalties based on device performance and behavior is essential.

\textbf{Resolution and Arbitration} Discrepancies and disputes are inevitable. A decentralized arbitration mechanism ensures that these are resolved fairly, with neutral arbiters possibly elected by the community.

\textbf{Periodic Audits \& Updates} Continuous improvement is vital. Regular audits, both of the software protocols and physical device checks, combined with protocol updates, ensure the network remains robust and up-to-date.

\subsection{Device Onboarding}

Device onboarding, the critical first step in the integration of a device into a Decentralized Physical Infrastructure Network (DePIN), sets the tone for the device's subsequent interactions and contributions to the network. 

Given the increasing vulnerabilities associated with hardware dependency, a software-centric approach is paramount. Below, we outline a formal process flow for software-based device onboarding in DePINs:

\begin{figure}[ht]
  \centering
  \includegraphics[scale=0.4]{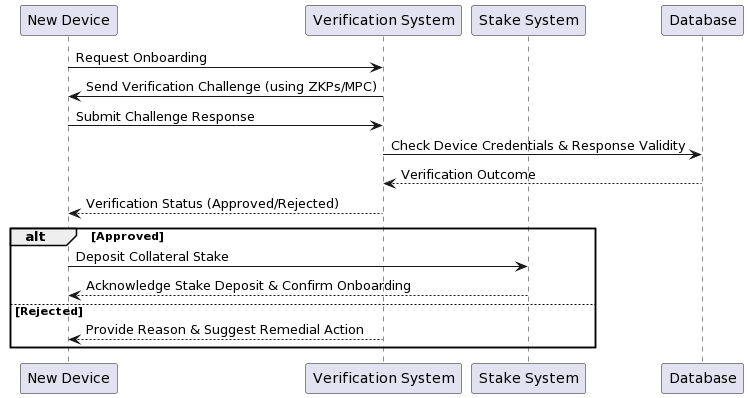}
  \caption{Sequence diagram of onboarding}
\end{figure}

\begin{enumerate}

\item \textbf{Device Registration Request} The device initiates the onboarding process by sending a registration request to the network. This request contains basic metadata about the device, including its type, model, and version, encrypted using the network's public key.

\item \textbf{Temporary Credential Assignment} The network assigns temporary credentials to the device, allowing it to participate in the initial stages of the onboarding process without granting full privileges.

\item \textbf{Zero-Knowledge Proof (ZKP) Validation}
      \begin{enumerate}
            \item The network challenges the device by sending a set of statements or assertions.
            \item Using ZKP, the device proves the veracity of these statements without revealing the actual data or computation1. This ensures device legitimacy without compromising sensitive information.
       \end{enumerate}
\item \textbf{Multi-Factor Authentication (MFA)} To bolster security, the device undergoes a multi-factor authentication process. This might involve time-based one-time passwords (TOTPs), cryptographic challenges, or other software-based authentication mechanisms.

\item \textbf{Device Behavior Analysis}The network employs machine learning models to analyze the device's behavior during the onboarding process. This step aims to detect anomalies or patterns inconsistent with legitimate devices.

\item \textbf{Credential Finalization}
\begin{enumerate}
\item  If the device successfully passes all verification stages, the network finalizes its credentials, replacing the temporary ones with permanent credentials.
\item These credentials are encrypted and sent to the device, ensuring secure communication for subsequent interactions.
\end{enumerate}

\item \textbf{Device Profile Creation}A unique profile for the device is created within the DePIN, storing essential details, including its public key, device type, and onboarding timestamp.

\item \textbf{Continuous Monitoring and Periodic Re-validation} Even after successful onboarding, the device's activities are continuously monitored. Periodic re-validation processes, leveraging techniques like ZKPs and MFA, ensure the device remains genuine and hasn't been compromised over time.

\item \textbf{Feedback Loop} After the onboarding process, feedback from the device is collected to improve and adapt the onboarding process continually. This feedback can encompass the device's experience during onboarding, challenges faced, and suggestions for improvement.

\end{enumerate}

\subsection{Data Transmission}

Within a Decentralized Physical Infrastructure Network (DePIN), the transmission of data is a cornerstone activity. Ensuring the integrity, authenticity, and confidentiality of this data flow is paramount for the network's overall health and functionality. 

One crucial mechanism that bolsters the trustworthiness of data transmission within DePINs is the integration of witnesses. In this section, we delve into the formal role and functioning of witnesses during data transmission in DePINs.

\begin{figure}[ht]
  \centering
  \includegraphics[scale=0.4]{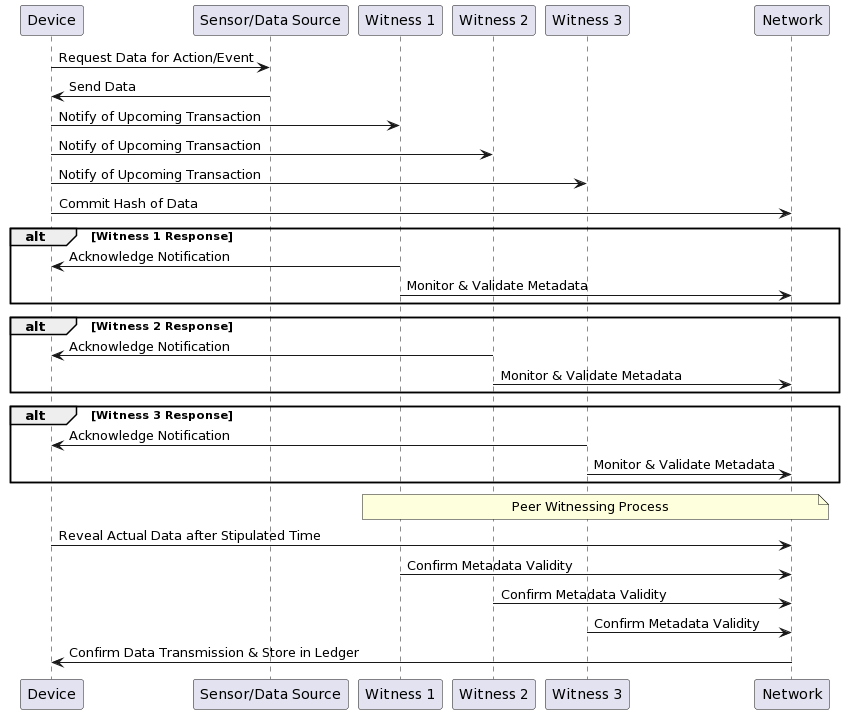}
  \caption{Sequence diagram of transmission}
\end{figure}

\textbf{Who are  Witnesses?} Witnesses, within the context of a DePIN, are neutral entities that observe, validate, and attest to the legitimacy of data transactions between participants. They are not directly involved in the transaction but serve to corroborate its authenticity and integrity.

\textbf{Witness Selection}
\begin{enumerate}
\item Witnesses can be randomly selected from the network based on predefined criteria such as their reputation, historical behavior, and computational capabilities.
\item To avoid collusion, the selection process ensures a diverse set of witnesses for each transaction, thereby minimizing the risk of coordinated malicious activities.
\end{enumerate}

\textbf{Data Validation by Witnesses}
\begin{enumerate}
\item When two or more participants engage in a data transaction, the involved data is encrypted and transmitted.
\item Selected witnesses receive a hashed version of this data or specific cryptographic challenges related to it.
\item Witnesses validate the data's integrity based on these hashes or challenges without directly accessing the raw data, preserving privacy.
\end{enumerate}

\textbf{Witness Attestation}
\begin{enumerate}
\item Upon successful validation, witnesses generate digital attestations or signatures confirming the data's authenticity.
\item These attestations are appended to the data transaction, serving as evidence of its validity.
\end{enumerate}

\textbf{Redundancy through Multiple Witnesses} Engaging multiple witnesses for a single transaction introduces redundancy. Even if one or a few witnesses are compromised or fail to validate the data, the presence of multiple attestations ensures a high degree of confidence in the transaction's legitimacy.

\textbf{Continuous Witness Evaluation} The network continuously evaluates the performance and behavior of witnesses. Witnesses that consistently provide accurate attestations are rewarded, while those showing anomalies or potential malicious behavior face penalties or expulsion.

\textbf{Resolving Conflicts}
\begin{enumerate}
\item In the rare event of conflicting attestations (where some witnesses validate a transaction while others reject it), a consensus mechanism comes into play.
\item The network re-evaluates the transaction, possibly engaging a fresh set of witnesses or using other validation techniques to determine the transaction's authenticity.
\end{enumerate}

\textbf{Feedback Loop with Witnesses} Just like devices, witnesses too can provide feedback on their experience, challenges faced during validation, or suggestions to improve the data transmission and validation process. This continuous feedback mechanism ensures that the witness system remains dynamic and adapts to emerging challenges.

Witnesses play a pivotal role in reinforcing the trustworthiness of data transmission within DePINs. By serving as neutral validators, they ensure that data transactions are not just taken at face value but undergo rigorous validation. This witness-based validation framework, combined with other security mechanisms, ensures that DePINs remain robust, transparent, and resistant to malicious activities.

\subsection{Consensus Mechanism}

In Decentralized Physical Infrastructure Networks (DePINs), the consensus mechanism forms the bedrock upon which data integrity and system synchronization are achieved. Given the decentralized nature of these networks, arriving at a unified agreement about the state of the ledger, especially in the context of data transmission, is vital. 

This section offers a formal exploration of the consensus mechanism, emphasizing its intricate relationship with data transmission.

\subsubsection{Need for Consensus} 

In a decentralized network, where multiple participants operate without a central authority, ensuring that all nodes agree on the validity and order of transactions is critical. This agreement prevents double-spending, data manipulation, and other potential fraudulent activities.

\textbf{Witness-Informed Consensus}
As discussed in the data transmission section, witnesses validate and attest to the authenticity of data transactions. These attestations serve as crucial inputs when nodes in the network determine the legitimacy of a transaction during the consensus process.

\textbf{Stages of Consensus Mechanism}
\begin{enumerate}

\item \textbf{Proposal Phase} A subset of nodes or participants proposes a set of transactions to be added to the ledger. This proposal includes the transaction data and, importantly, the attestations from witnesses.

\item \textbf{Validation Phase} Other nodes in the network validate the proposed transactions. This involves checking witness attestations, ensuring that the transaction adheres to network rules, and confirming that there are no conflicting transactions.

\item \textbf{Commitment Phase} Once a majority (or a predefined threshold) of nodes agree on the validity of the proposed transactions, they are added to the ledger. The network then moves to the next set of transactions.
\end{enumerate}

\textbf{Handling Conflicts} In scenarios where there's disagreement among nodes (e.g., due to conflicting witness attestations), the consensus mechanism initiates additional validation rounds. Fresh witnesses might be engaged, or other conflict resolution strategies, such as weighted voting based on stake or reputation, might be employed.

\textbf{Continuous Synchronization} To ensure that all nodes maintain an identical copy of the ledger, continuous synchronization occurs post-consensus. Nodes share and update their ledgers to reflect the latest agreed-upon state.

\subsection{Anomaly Detection}

In the intricate web of Decentralized Physical Infrastructure Networks (DePINs), the constant flow of data and transactions demands rigorous oversight. Beyond traditional validation and consensus mechanisms, a proactive approach is essential to identify and address potential aberrations. 

Anomaly detection, a sophisticated method of discerning patterns that do not conform to expected behavior, serves as a sentinel for these networks, flagging unusual activities that might indicate malicious intent, system faults, or data inconsistencies.

\textbf{Significance of Anomaly Detection}
 In a decentralized network, where myriad devices and nodes interact, the sheer volume and complexity of data can mask unauthorized or malicious activities. Anomaly detection, by continually analyzing this data, acts as a vigilant guardian, ensuring that even subtle deviations do not go unnoticed.

\textbf{Mechanisms of Anomaly Detection}

\begin{enumerate}
\item \textbf{Statistical Models} These models establish a baseline of 'normal' behavior based on historical data. Any deviation from this established norm, be it sudden spikes in transaction volume or unusual data packet sizes, can be flagged for further investigation.

\item \textbf{Machine Learning Models} Machine learning algorithms, especially unsupervised ones, can be trained on vast datasets to recognize patterns. Once trained, they can identify anomalies by spotting data points or patterns that deviate from the learned norm.

\item \textbf{Time-Series Analysis}
 Given that many data transactions in DePINs are sequential, time-series analysis can detect anomalies by examining data points in their chronological context, highlighting inconsistencies or abrupt changes over time.

\end{enumerate}

\textbf{Integration with Data Transmission and Consensus Mechanism}
Anomaly detection operates in tandem with data transmission and the consensus mechanism. As data is transmitted and validated, the detection systems analyze it in real-time, ensuring that anomalies are identified and addressed before they can impact the network's integrity.

\textbf{Response to Detected Anomalies}

\begin{enumerate}

\item \textbf{Immediate Alerts}
Upon detecting an anomaly, the system triggers immediate alerts to designated nodes or administrators, ensuring rapid response.

\item \textbf{Quarantine Procedures}
Potentially malicious or compromised nodes or data packets can be temporarily isolated, preventing any potential harm to the broader network.

\item \textbf{Investigation and Resolution}
Detailed logs of the anomalous activity are examined to determine the root cause, be it system faults, external attacks, or data inconsistencies. Appropriate corrective actions, ranging from patching vulnerabilities to reinforcing security protocols, are then initiated.

\end{enumerate}

\textbf{Continuous Learning and Adaptability}
An effective anomaly detection system is not static. It continuously learns from new data, refining its models and adapting to the evolving behavior of the network. This ensures that it remains effective even as the network grows and changes.

\subsection{Rewards \& Penalties}

\textbf{Rationale for a Dual Mechanism}
Given the heterogeneous nature of participants in a DePIN, a one-size-fits-all approach is insufficient. While rewards motivate and encourage nodes and devices to act in the network's best interest, penalties serve as a deterrent against potential malicious intent or negligence.

\textbf{Reward Mechanisms}
\begin{enumerate}

\item \textbf{Performance-Based Incentives}
Nodes or devices that consistently demonstrate honest behavior, validate transactions accurately, or contribute positively to the network's health receive performance-based rewards. This can be in the form of cryptocurrency tokens, increased reputation scores, or preferential treatment in transaction processing.

\item \textbf{Contribution-Based Rewards}
Nodes that offer additional resources, be it computational power, storage, or bandwidth, receive proportional rewards. This incentivizes participants to contribute more to the network's robustness and efficiency.

\item \textbf{Longevity and Loyalty Bonuses}
Nodes or devices that have been part of the network for an extended period and have maintained a positive track record receive bonuses. This encourages long-term commitment and stability within the network.
\end{enumerate}

\textbf{Penalty Mechanisms}
\begin{enumerate}

\item \textbf{Stake Forfeiture}
Many DePINs operate on a staking model where nodes deposit collateral to participate. Malicious activity or non-compliance can result in a partial or complete forfeiture of this stake, serving as a strong financial deterrent.

\item \textbf{Reputation Degradation}
Nodes or devices found to be acting against the network's interests experience a reduction in their reputation score. A degraded reputation can result in reduced rewards, slower transaction processing, or even exclusion from certain network activities.

\item \textbf{Temporary or Permanent Bans}
Severe violations can result in nodes or devices being temporarily banned from participating in the network. In cases of repeated offenses or extremely malicious intent, permanent expulsion might be enforced.
\end{enumerate}

\textbf{Transparent Evaluation Metrics}
For the rewards and penalty mechanism to be effective and fair, the criteria for evaluations must be transparent and known to all participants. This includes clear metrics on what constitutes positive contributions, the thresholds for penalties, and the means of redressal in case of disputes.

\textbf{Continuous Review and Adaptation}
The rewards and penalty mechanisms are not static. They undergo continuous review based on network feedback, emerging challenges, and technological advancements. This ensures that the system remains fair, effective, and relevant in the face of evolving network dynamics.

\subsection{Resolution and Arbitration}

In the intricate landscape of Decentralized Physical Infrastructure Networks (DePINs), the absence of centralized governance brings forth the inevitability of disputes and discrepancies. Addressing these conflicts in a manner that upholds the principles of fairness, transparency, and efficiency becomes paramount. This section delves into the academic intricacies of the resolution and arbitration mechanisms tailored for DePINs, ensuring that conflicts are managed and resolved with the utmost rigor.

\begin{figure}[ht]
  \centering
  \includegraphics[scale=0.4]{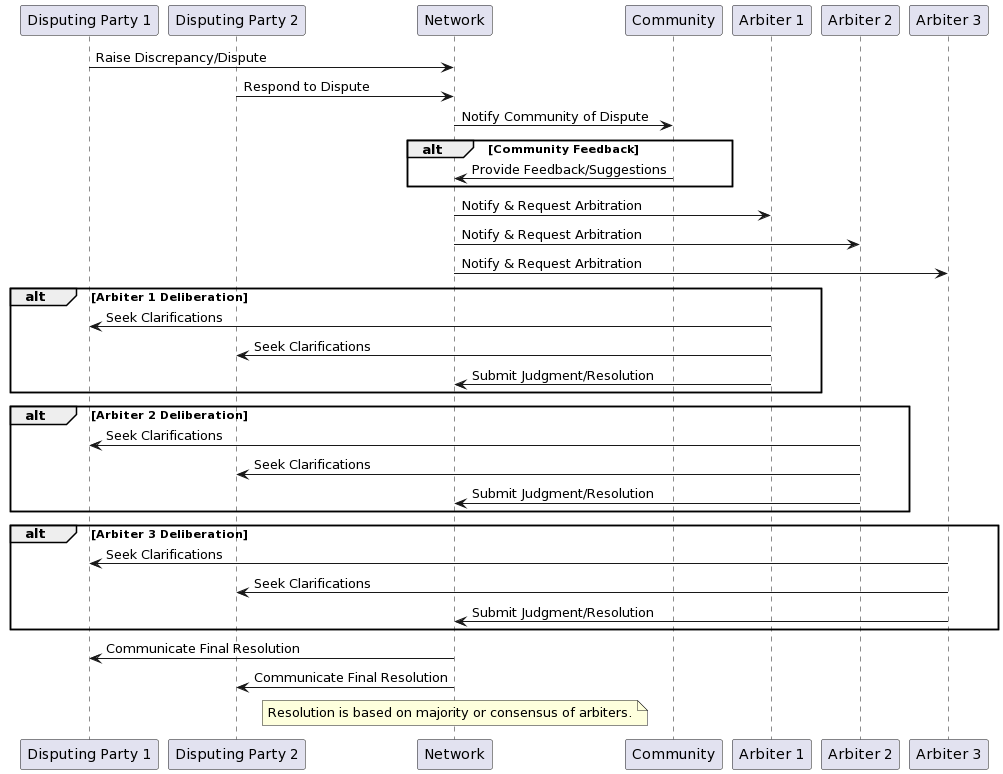}
  \caption{Sequence diagram of arbitration}
\end{figure}

\textbf{Decentralized Resolution}
Centralized conflict resolution mechanisms, while streamlined, may not align with the decentralized ethos of DePINs. They can introduce single points of failure, potential biases, or even avenues for manipulation. A decentralized approach to resolution and arbitration ensures that decisions are derived from a broader, more representative subset of the network, maintaining the trust and integrity inherent to DePINs.

\textbf{Stages of Decentralized Resolution and Arbitration}

\begin{enumerate}

\item \textbf{Initial Mediation}
When a dispute arises, initial mediation is facilitated by algorithmically-selected mediators from the network, chosen based on criteria like reputation, expertise in the subject of dispute, and historical conflict resolution performance. This stage aims for a swift, amicable resolution without escalating the dispute.

\item  \textbf{Community Review}
If mediation fails, the dispute is presented to a broader segment of the network. Participants can review the details, ensuring transparency, and collectively weigh in, leveraging mechanisms like weighted voting based on stake or reputation.

\item  \textbf{Arbitration Panel Selection}
Should the community review not yield a resolution, an arbitration panel is stochastically selected from a pool of vetted arbitrators within the network. These arbitrators possess specialized knowledge, experience, and have undergone training to handle complex disputes.

\item  \textbf{Final Arbitration}
The arbitration panel reviews the dispute in-depth, considering all evidence, testimonials, and network data. Their decision, once reached, is binding and is recorded on the blockchain for transparency and immutability.

\end{enumerate}

\textbf{Redressal and Appeal Mechanisms}
Even within a decentralized system, the possibility of oversights or biases, albeit reduced, remains. An appeal mechanism allows parties to challenge the decision of the arbitration panel, but with stringent criteria to prevent frivolous appeals. A separate set of arbitrators, distinct from the initial panel, reviews appeal cases.

\textbf{Integration with Reward and Penalty Systems}
The outcomes of resolutions often tie back to the network's reward and penalty mechanisms. For instance, a party found at fault might face penalties, while whistleblowers or those wrongly accused may receive compensations or reputation boosts.

\textbf{Continuous Training and Feedback}
Arbitrators, given their pivotal role, undergo continuous training sessions, staying updated with the latest network protocols, ethical guidelines, and conflict resolution techniques. Regular feedback loops ensure that the arbitration process is continually refined, adapting to the evolving dynamics of the network.

\subsection{Stochastic checks}

Stochastic checks, or random inspections, serve as an integral layer of security within Decentralized Physical Infrastructure Networks (DePINs). Their unpredictable nature ensures that participants maintain genuine behavior, as they remain unaware of when they might be inspected. 

Integrating stochastic checks into the consensus mechanism and data transmission processes further fortifies the system against potential threats and malicious activities.

\subsubsection{Data Transmission}
During the process of data transmission, not every transaction might be inspected in detail due to the sheer volume. However, a random subset of these transactions undergoes detailed stochastic checks.
\begin{enumerate}

\item \textbf{Random Witness Selection} While witnesses are already part of the data transmission process, a random subset of these witnesses can be chosen stochastically for deeper transaction scrutiny.

\item \textbf{Random Packet Analysis} Periodically, certain transmitted data packets are chosen at random for a deep-dive analysis, ensuring that even routine transactions are occasionally subjected to intense scrutiny.
\end{enumerate}

\subsubsection{Consensus Mechanism}

\begin{enumerate}

\item \textbf{Proposal Phase with Random Challenges}
During the proposal phase, nodes or participants proposing a set of transactions might be stochastically presented with random challenges or puzzles to solve. This ensures that the proposing nodes are genuine and prevents Sybil attacks.

\item \textbf{Validation Phase with Random Validators}
Instead of a fixed set of validators, a random subset of nodes is chosen stochastically in each validation phase. This unpredictability makes it harder for malicious nodes to collude or anticipate which nodes will validate their transactions.

\item \textbf{Commitment Phase with Random Delays}
Post validation, before adding transactions to the ledger, random delays might be introduced stochastically. These delays can serve as windows during which any node can raise concerns or flag potential anomalies in the proposed transactions.

\end{enumerate}

\subsubsection{Continuous Synchronization}
As nodes synchronize their ledgers post-consensus, stochastic checks can randomly verify the integrity of the data being synchronized, ensuring that nodes aren't propagating altered or malicious versions of the ledger.

\subsubsection{Further integrations}
Stochastic checks can also be integrated into other processes like device onboarding, where random devices undergo detailed inspections, or during witness evaluations, where witnesses might be stochastically challenged to prove their legitimacy

\section{Conclusion}

The exploration and development of decentralized systems have experienced significant attention in contemporary academic discourse. Within this landscape, the Generalised DePIN Protocol emerges as a pivotal contribution, addressing the intricate challenges associated with Decentralized Physical Infrastructure Networks. This concluding section seeks to delineate the overarching implications of the GDP, highlighting its academic and practical significance.

Central to the GDP's architecture is its meticulous approach to device onboarding, emphasizing software-driven techniques. This emphasis not only mitigates potential vulnerabilities associated with hardware manufacturer dependencies but also charts a course for enhancing system-wide trust and security. The delineated process flows and the integration of witnesses in data transmission underscore the protocol's commitment to operational reliability.

Moreover, the GDP's anomaly detection mechanisms, underpinned by rigorous mathematical and computational methodologies, offer robust safeguards against potential system abuses. When juxtaposed with the protocol's carefully calibrated rewards and penalty system, it becomes evident that the GDP is designed to foster a culture of integrity and transparency among network participants.

The protocol's approach to decentralized conflict resolution, grounded in established arbitration principles, further accentuates its academic rigor. By facilitating a decentralized mechanism for dispute management, the GDP introduces an innovative paradigm that holds potential for broader applications in decentralized systems governance.

In evaluating the broader implications of the GDP, one cannot overlook its potential to serve as a foundational framework for various industry applications. From decentralized energy grids to ridesharing platforms, the principles encapsulated within the GDP offer transformative possibilities that transcend traditional centralized infrastructures.

In summation, the Generalised DePIN Protocol stands as an academically robust and practically relevant contribution to the domain of decentralized systems. Its methodical architecture, combined with its forward-looking provisions, position the GDP as an instrumental tool for future research and applications. The protocol's implications for Decentralized Physical Infrastructure Networks underscore the evolving dynamics of decentralized systems, offering insights and methodologies that will undoubtedly influence subsequent academic and industry endeavors.

\bibliographystyle{abbrvnat}
\bibliography{sample}
\end{document}